\documentclass[aps,prb,reprint,showpacs,superscriptaddress,groupedaddress]{revtex4-1}
\usepackage{graphicx}
\usepackage{amsmath}
\usepackage{amssymb}
\usepackage{dcolumn}
\usepackage{latexsym}
\usepackage{rotating}
\usepackage[usenames,dvipsnames]{xcolor}
\usepackage{float}
\usepackage{epsfig}
\usepackage{psfrag}
\usepackage{natbib}
\usepackage{bm}
\usepackage{eucal}
\usepackage{braket}
\usepackage{enumerate}
\usepackage{longtable}
\usepackage{subfigure}
\usepackage{bm}
\usepackage{hyperref}
\usepackage{amsfonts}
\usepackage{dcolumn}% Align table columns on decimal point
\usepackage{bm}
\usepackage{subfigure}

\newcommand{\be}{\begin{equation}}
\newcommand{\ee}{\end{equation}}
\newcommand{\bn}{\begin{eqnarray}}
\newcommand{\en}{\end{eqnarray}}

\usepackage{color} % use colored text in latex

\usepackage{hyperref}
\begin{document}

\author{S. Koley$^{1}$}
\thanks{Present address: Department of Physics, Amity Institute of Applied Sciences, Amity University, Kolkata, W.B. 700135, India.}
\author{N. Mohanta$^{2}$}
\thanks{Present address: Material Science and Technology Division, Oak Ridge National Laboratory, Oak Ridge, TN 37831, USA.}
\author{A. Taraphder$^{3}$}

\title{Charge Density Wave and Superconductivity in Transition Metal Dichalcogenides}
\affiliation{$^{1}$Department of Physics, North Eastern Hill University,
Shillong, Meghalaya, 793022 India.\\
$^{2}$Department of Physics, Indian Institute of Technology Kharagpur, W.B. 721302, India\\
$^{3}$Department of Physics and Centre for Theoretical Studies, Indian Institute of Technology Kharagpur, W.B. 721302, India
}

\begin{abstract}
Competing orders in condensed matter give rise to the emergence of fascinating, new phenomena. Here, we investigate the competition between superconductivity and charge density wave in the context of layered-metallic compounds, transition metal dichalcogenides, in which the superconducting state is usually suppressed by the charge density wave. We show, using real-space self-consistent Bogoliubov-de Gennes calculations and momentum-space calculations involving  density-functional theory and dynamical mean-field theory, that there is a surprising reappearance of superconductivity in the presence of non-magnetic disorder fluctuations, as observed in recent experiments.  
\end{abstract}
\maketitle
\section{Introduction}

\noindent The issue of disorder and Coulomb interaction, competing with each other, is an old one. 
In the absence of interaction, disorder is known to localize electronic states. The 
problem appears in systems where superconductivity (SC) competes with 
long-range orders like charge density wave (CDW)~\cite{at1,sao,kim} and spin 
density wave (SDW)~\cite{miz}. There is a large body of literature on the 
competition between SC and SDW in presence of disorder~\cite{allou}, while 
much less is discussed on SC and CDW competing with each other.

\noindent In the weak coupling case, these orders compete for the same Fermi surface and 
when disorder suppresses one, the other grows at its expense. Note that for 
s-wave SC, non-magnetic disorder (unless very strong) has little effect, while 
it pushes the CDW instability down. From a traditional view point, 
superconductivity and CDW are considered as weak coupling Fermi surface 
instabilities mediated by phonons~\cite{grun}. There have been reports of both 
cooperation~\cite{kiss} and competition~\cite{boris1,at1} between the two, 
though on the face of it, they compete for the same electrons. The original idea
 of a CDW due to Fermi-surface nesting (Peierls instability) has rarely been 
borne out in real systems where nesting is scarce. Moreover, even if it exists, 
nesting appears at the wrong wave vectors and away from the Fermi surface quite 
often~\cite{mazin,boris1}. Therefore, such single-particle mean-field pictures, 
invoking large segments of the Fermi surface nested with each other, need to be 
abandoned in favor of an alternative, strong coupling view. The strong coupling 
approach is further supported by the finding of $2\Delta/k_BT_{_{SC}} > 5$ (where $\Delta$ is the 
superconducting pairing gap at zero temperature and $T_{_{SC}}$ is the superconducting transition temperature) in many transition metal dichalcogenides (TMDs)~\cite{mcm} (about 7 for 1T-TiSe$_2$ and 10 for 2H-TaSe$_2$). 
 
\noindent The TMDs have typically shown bad metallic resistivity above  $T_{_{CDW}}$, the CDW ordering temperature,
becoming quadratic and more metallic below it. There is often little or no 
anomally across the transition. The linear resistivity at high temperature is 
persistent till fairly high temperatures~\cite{dord, vesco}. In spite of such 
overwhelming indications to the contrary, and the near-absence of required 
Fermi-surface nesting, much of the theoretical work on TMDs took recourse to 
mean-field considerations. Mottness has not been considered important for TMDs, 
except for 1T-TaS$_2$~\cite{fazek} long back and for 2H-TaSe$_2$ ~\cite{at2,vesco}
 and 1T-Tise$_2$ ~\cite{at3} lately. Large spectral weight transfer and absence 
of phonon signatures at typical energies in ARPES data, along with 
momentum-independedent self-energy~\cite{valla} are strong pointers to physics 
beyond (mean-field) single-particle processes.

\noindent Superconductivity in most of the TMDs also posseses interesting questions. It 
usually appears when the CDW is suppressed by disorder, intercalation or 
pressure~\cite{kusmart}. One such approach has recently been taken by Li, et al.,
~\cite{li_npj} where, in 2H-TaSe$_2$, Se is replaced by sulphur (S) gradually, 
all the way to 2H-TaS$_2$. They report an emergence of robust superconducting 
order in single crystal TaSe$_{2-x}$S$_x$ ($0<x<2$) alloy. The SC transition 
temperature (T$_{SC}$) of the alloy is more than that of the two end-members TaSe$_2$ and TaS$_2$. The conductivity near the middle of the alloy series is also found to be higher than that of either 2H-TaSe$_2$ and 2H-TaS$_2$. This observation clearly suggests that superconductivity in this system is in competition with CDW.  

\noindent In order to understand this competition, we take up the problem of co-existing 
CDW and SC and add non-magnetic disorder to it. Two kinds of disorder are studied: random 
disorder (uniform probabability in real space with fixed strength) and clusters 
of disordered regions of fixed strengths. Using a self-consistent Bogoliubov-de-Gennes (BdG) 
formulation, we observe the evolution of SC and charge order using a Monte Carlo
 method~\cite{narayan_jpcm}. Such work has been used in the context of disordered
 superconductivity during the heyday of high T$_c$ 
superconductivity~\cite{daggo,bc}. 

\noindent In addition, we also approach the alloy problem from a strong-coupling point of 
view at specific dopings. A density functional theory, followed by multi-orbital
 dynamical mean field theory (DFT+MO-DMFT) calculations for TaSe$_{2-x}$S$_{x}$ (at $x$=0 and 1) are performed to undesrtand the nature of the competition 
between CDW and SC. The disorder-induced competition is introduced via doping of sulfer which affects the band structure slightly, while acting as a substitutional disorder. The strong-coupling approach views the CDW 
as a condensate out of a pre-formed excitonic liquid~\cite{at2,at3} of bad 
metals. The unconventional CDW found in the parent dichalcogenide is a 
bose-condensate of pre-formed excitons from high temperature. As there is no 
magnetism found 
in TMD, the alloy  is a bad metal without magnetic fluctuations. Therefore 
direct comparisons could be drawn with non-magnetic disorder that  degrades CDW 
and facilitates SC at low temperatures. The paper is arranged as follows: we discuss the methods employed in the following section. We then discuss the results and draw our conclusions at the end. 

\section{Methods}
\noindent To investigate the interplay between charge density wave and superconductivity 
in the presence of real-space potential fluctuations, we use a self-consistent 
BdG formalism in which the s-wave superconducting pairing gap ($\Delta_{_{SC}}$)
and charge density wave order parameter ($\Delta_{_{CDW}}$) are computed for a given disorder configuration. \\

\noindent \textit{Self-consistent BdG formalism:} 
In this method, we consider a square lattice of size $N= 41\times 41$, with open boundary conditions, and the order parameters, $\Delta_{_{SC}}$ and $\Delta_{_{CDW}}$, are computed self-consistently, until convergence is achieved at each lattice site. 

\begin{figure*}[t]
\begin{center}
\epsfig{file=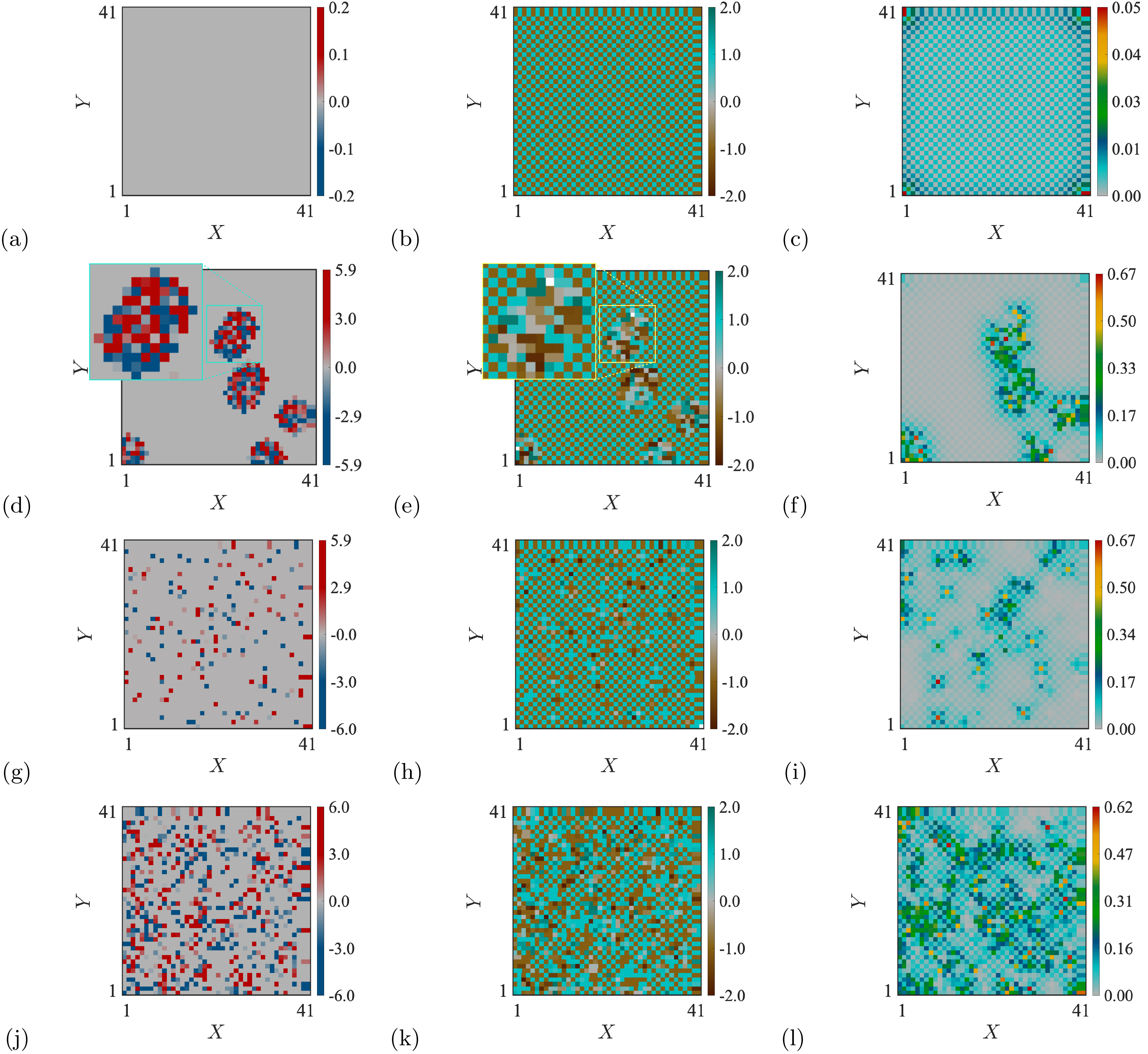,trim=0.0in 0.0in 0.0in 0.0in,clip=false, width=160mm}
\caption{(Color Online) Spatial profiles of disorder (first column), CDW order (represented by carrier density modulation $\xi_i$) (second column) and superconducting pairing amplitude (third column) for a configuration of clustered disorder at strength $V_d$ =0 ((a)-(c)), $V_d$ =6 ((d)-(f)). The same profiles with random disorder at a combination of $V_d$=6 and concentration $10\%$ ((g)-(i)) and  and $40\%$ ((j)-(l)).}
\label{fig1}
\end{center}
\end{figure*}

\noindent We  consider  the  following  BdG  Hamiltonian for  disorder-affected
superconductor  on a  square lattice
\begin{align}
{\cal H}_{BdG}=&-t\sum_{\langle ij \rangle,\sigma}(c_{i\sigma}^\dagger c_{j\sigma}+H.c.)-
\sum_{i,\sigma}(\mu-V{_i}) c_{i\sigma}^\dagger c_{i\sigma} \nonumber \\
&+\sum_{i}(\Delta_{_{SC}}^ic_{i\uparrow}^{\dagger}c_{i\downarrow}^{\dagger}+H.c.)
\label{Hbdg}
\end{align}
\noindent where $t$ is the nearest-neighbor electron hopping energy, $V_i$ is 
the local non-magnetic disorder in the chemical potential $\mu$. Two types of configuration of the onsite disorder $V_i$ have been considered in this study: 
(i) clustered regions of the disorder as shown in Fig.\ref{fig1}d and (ii) 
randomly distributed disorder as shown in Fig.\ref{fig1}g and Fig.\ref{fig1}f. 
The onsite disorder $V_i$ is varied within a range $-V_d\leq V_i \leq V_d$, where $V_d$ is the strength of the disorder.  ${\cal H}_{BdG}$ is diagonalized  via a Bogoliubov-Valatin transformation
$\hat{c}_{i\sigma}=\sum_{i,\sigma^{\prime}}u_{n\sigma\sigma^{\prime}}^i\hat{\gamma}_{n\sigma^{\prime}}+v_{n\sigma\sigma^{\prime}}^{i*}\hat{\gamma}^{\dagger}_{n\sigma^{\prime}}$
which gives the following equation for the local superconducting order parameter
$\Delta_{_{SC}}^i=-U_{_{SC}} \langle c_{i\uparrow}c_{i\downarrow} \rangle$, where $U_{_{SC}}$ is the strength of the pairwise electron-electron attractive interaction,  in   terms   of   the
Bogoliubov quasiparticle and quasihole amplitudes $u_{n\sigma}^i$ and $v_{n\sigma}^i$
\begin{equation}
\begin{split}
\Delta_{_{SC}}^i=&-U_{_{SC}}\sum_{n}[u_{n\uparrow}^iv^{i*}_{n\downarrow}(1-f(E_n))
+u_{n\downarrow}^i v^{i*}_{n\uparrow}f(E_n)]
\end{split}
\label{delsi}
\end{equation}
\noindent  where $f(E_n)=1/(1+e^{E_n/{k_BT}})$  is the  Fermi  function at
temperature  $T$ and at the $n^{\rm th}$ energy eigenvalue $E_n$. 
The quasiparticle and quasihole amplitudes $u_{n\sigma}^i$ and
$v_{n\sigma}^i$  are  determined   by  solving  the  following  BdG
equations
\begin{equation}
{\cal H}_{BdG}~\phi_n^i=\epsilon_n\phi_n^i
\label{bdg_eq}
\end{equation}
written in the basis $\phi_n=[u_{n\uparrow}^i,u_{n\downarrow}^i,v_{n\uparrow}^i,v_{n\downarrow}^i]$.

\noindent The above equations (Eq.~\ref{delsi} and Eq.~\ref{bdg_eq}) are solved self-consistently on a finite, two dimensional square lattice with open boundary conditions and the superconducting order parameter is determined using $\Delta_{_{SC}}=\frac{1}{N}\sum_i\Delta_{_{SC}}^i$, averaged over several realizations of disorder configuration. 

\noindent The CDW order is modeled via a real-space modulation of the onsite potential, expressed by the following Hamiltonian, which is added to Eq.~\ref{Hbdg} for the self-consistent solution,
\begin{equation}
H_{_{CDW}} = -U_{_{CDW}} \sum_{i} ( \cos(Q_x.x + \alpha)+\cos(Q_y.y + \beta) ) c_{i}^{\dagger} c_{i},
\end{equation}
where $U_{_{CDW}}$ defines the strength of the CDW order, $Q_x$, $Q_y$ are the CDW wave vectors, $\alpha$ and $\beta$, when non-zero, give incommensurate CDW.  The local CDW order parameter is calculated using $\Delta_{_{CDW}}^i=(1/U_{_{CDW}})(2n- \xi_{i})$, where $n=(1/N) \sum_{i}\langle c_{i\sigma}^{\dagger} c_{i\sigma} \rangle$ is the total average carrier density at a given set of parameters and $\xi_{i}$ is the local carrier density due to CDW. The mean CDW order parameter is obtained, as in the previous case of SC order parameter, by averaging over all lattice sites $\Delta_{_{CDW}}=(1/N) \sum_{i}\Delta_{_{CDW}}^i$. In the present study, we consider CDW order to be dominant over the SC order in the homogeneous situation ($V_d=0$) and choose $U_{_{SC}}=2t$ and $U_{_{CDW}}=4t$ throughout the manuscript, with no qualitative change in the result for other choices, as long as $U_{_{CDW}}>U_{_{SC}}$. Also, we express all energies for the results present here, obtained from the BdG analysis, in units of the hopping energy scale $t$ as wet set $t=1$. \\

\noindent \textit{Combined DFT+DMFT approach:} 
2H-TaSe$_2$ has a layered hexagonal structure (space group P63/mmc). The 
crystal structure contains layers of transition metals and chalcogens, aranged 
along the z direction; these layers are separated by van der Waals gap. First 
principles calculations were performed using WIEN2k ~\cite{wien2k} full-potential
linearized augmented plane wave (FP-LAPW) ab initio package within the DFT 
formalism to get the band structure, density of states (DOS) and related 
quantities. A $10\times 10\times 10$ k-mesh (with the cutoff parameter, 
Rk$_{max}$ = 7.5) is used here and a generalized gradient approximation 
Perdew-Burke-Ernzerhof (GGA-PBE) exchange correlation potential is chosen. 
The muffin-tin radius, R$_{mt}$ is chosen to be 2.5 a.u. for Ta, 2.39 for Se, and 2.21 for S (used for doping). The cell parameters are
derived from an earlier experiment~\cite{li_npj}. Finally the self-consistent field (scf) calculations are evaluated with an accuracy of 0.0001 eV and the energy-minimized crystal structure is obtained for the parent and doped dichalcogenide. From the converged structure we determined the band structure and DOS. To carry out transport calculations, a fully charge-self-consistent dynamical mean field theory (DMFT) is used. In a strongly-correlated system, like a TMD, DFT, combined with DMFT, has been successful in explaining a host of experimental results. The multi-orbital iterated perturbation theory (MO-IPT), a computationally fast and effective impurity solver, has been used here. Though not exact, it works nicely in real systems for both high and low temperature regimes. Here we have used multi-orbital Hubbard model with reasonable intra and inter-orbital Coulomb interactions. The total Hamiltonian is expressed as,
%\begin{equation}
%H=\sum_{k,a,b}t_{k,a,b}c^{\dagger}_{k,a}c_{k,b}+U\sum_{i,a}n_{ia\uparrow}n_{ia\downarrow} + $$
%$$U'\sum_{i,a,b}n_{ia}n_{ib}
%\end{equation}
\begin{align}
H=\sum_{k,a,b}t_{k,a,b}c^{\dagger}_{k,a}c_{k,b}+U\sum_{i,a}n_{ia\uparrow}n_{ia\downarrow}
+U'\sum_{i,a,b}n_{ia}n_{ib}
\end{align}
where $a$, $b$ denote the DFT conduction (Ta-d) band and valence (Se-p and S-p) bands) with dispersions $t_{aa} , t_{bb}$ (calculated from WIEN2K pogram). $t_{ab}$ is the interorbital hopping and $U$ and $U'$ are the 
intra- and inter-orbital Coulomb repulsions. Initially $t_{ab}$ is taken as 0.4 
eV for 2H-TaSe$_2$ which is effectively modified inside DMFT self consistency 
equation according to the modified band structure and doping. $U$ and $U'$ are 
chosen as 1.0 eV and 0.5 eV after checking for the possible values to show the 
CDW induced changes in resistivity.
%---------------------------------------------
%---------------------------------------------
%---------------------------------------------
\begin{figure*}[t]
\begin{center}
\epsfig{file=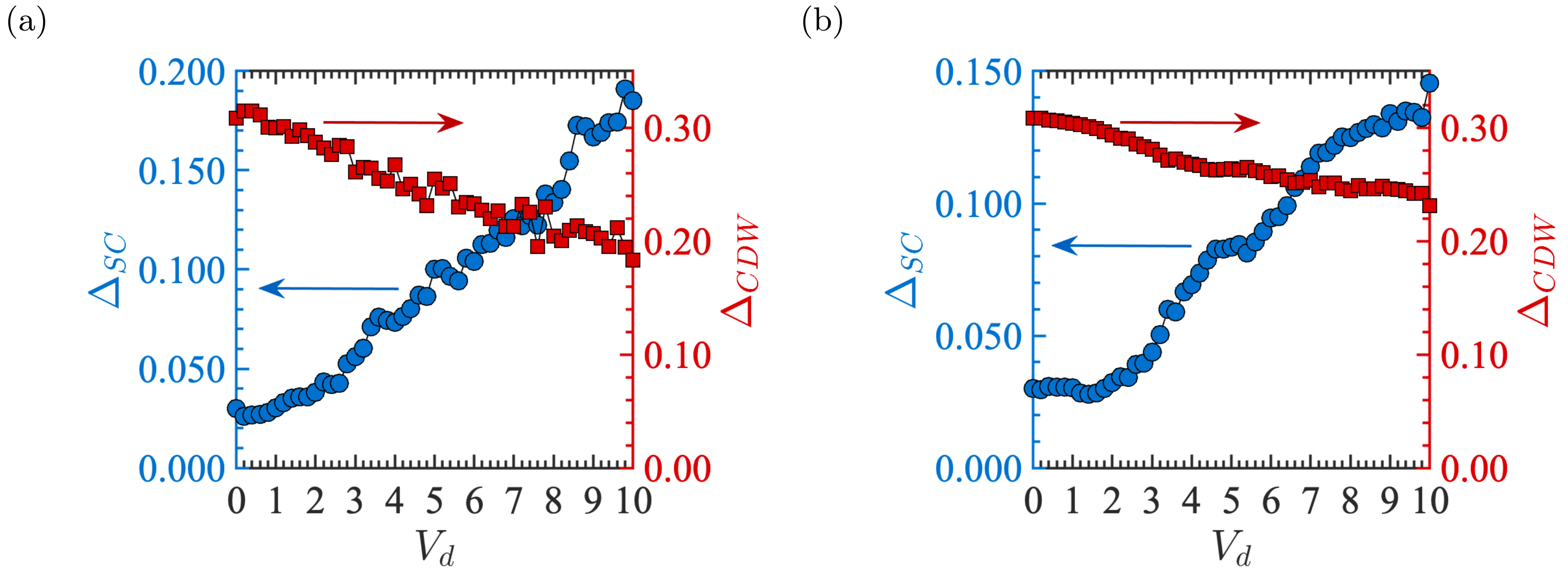,trim=0.0in 0.0in 0.0in 0.0in,clip=false, width=160mm} 
\caption{(Color Online) Variation of $\Delta_{CDW}$ and $\Delta_{SC}$ with 
disorder strength in (a) clustered disorder configuration and (b) random 
disorder configuration. In clustered disorder $\Delta_{CDW}$ decreases 
with disorder strength and it follows the same trend in random disorder. 
With both disorder at a concentration strength $V_d$ = 7  both superconductivity 
and CDW coexists and superconducting order parameter increases with 
increasing $V_d$.}
\label{fig2}
\end{center}
\end{figure*}
%---------------------------------------------
%---------------------------------------------
\begin{figure}[h]
\begin{center}
\epsfig{file=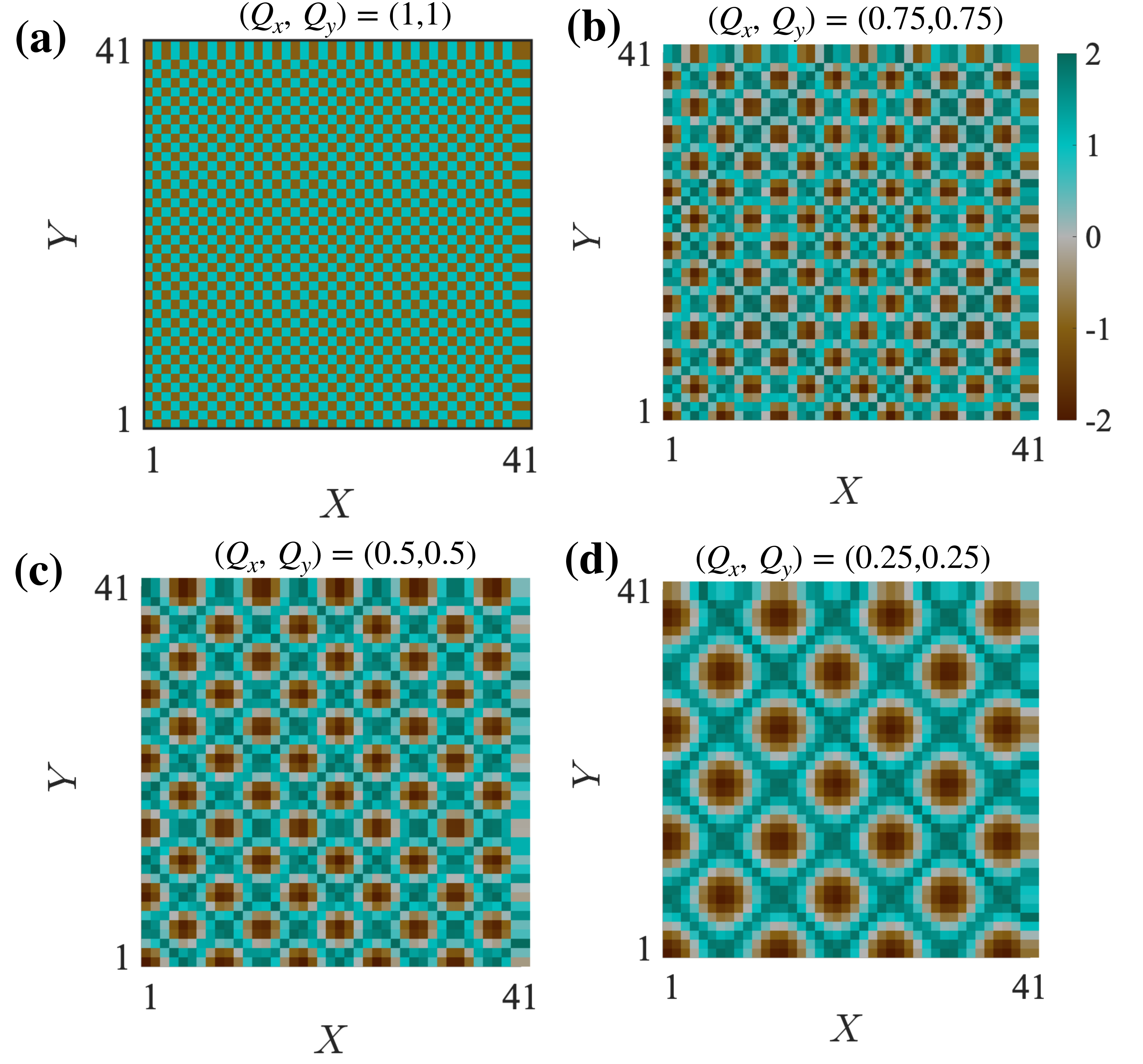,trim=0.0in 0.0in 0.0in 0.0in,clip=false, width=90mm} 
\caption{(Color Online) Spatial profile of the carrier density modulation $\xi_i$, revealing the CDW pattern at different Q vectors (a) ($Q_x$, $Q_y$)=(1,1), (b) ($Q_x$, $Q_y$)=(0.75,0.75), (c) ($Q_x$, $Q_y$)=(0.5,0.5) and (a) ($Q_x$, $Q_y$)=(0.25,0.25).}
\label{fig4}
\end{center}
\end{figure}
%---------------------------------------------
%---------------------------------------------
\begin{figure*}[t]
\begin{center}
\epsfig{file=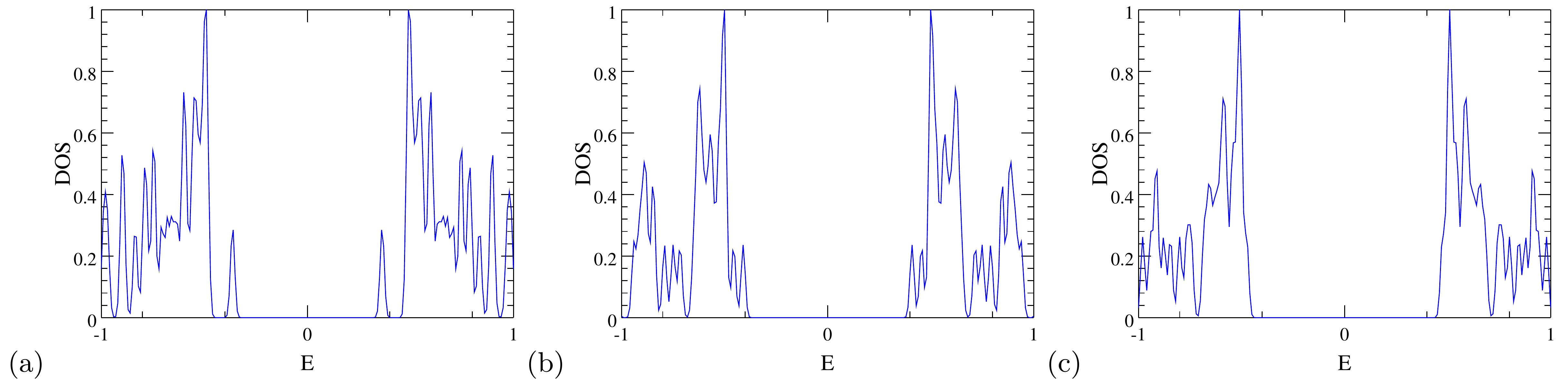,trim=0.0in 0.0in 0.0in 0.0in,clip=false, width=160mm} 
\caption{(Color Online) Density of states averaged over all sites and disorder 
at different clustered disorder strength (a) 2, (b) 6, (c)10 and disorder 
concentration $40\%$. spectral gap is evident throughout all the disorder 
strength and it increases with the strength.}
\label{fig3}
\end{center}
\end{figure*}
%---------------------------------------------
\section{RESULTS}
\noindent \textit{Results from self-consistent BdG calculations:} We present our results from the BdG calculation on an extended Hubbard model first. In our model, we have considered a 2D $s$-wave superconductor with CDW order and non-magnetic disorder. With the parameters (unit of energy $t=1$) mentioned above we averaged all the observables for 200 independent realizations of two different types of disorder: (i) a random choice of disorder sites without any bias and (ii) a clustered disorder configuration.

\begin{figure*}[h]
\begin{center}
\epsfig{file=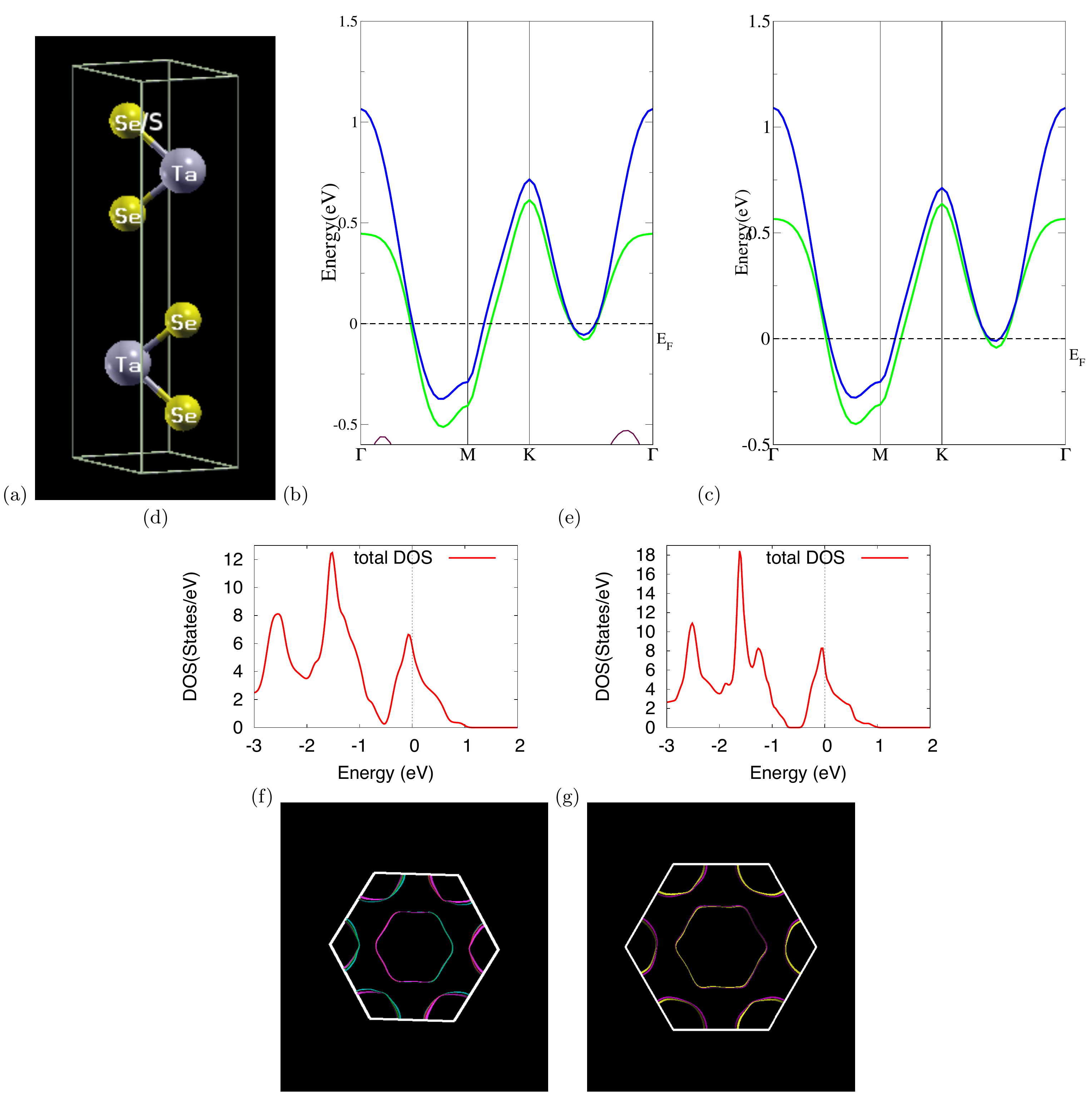,trim=0.0in 0.0in 0.0in 0.0in,clip=false, width=160mm} 
\caption{(Color Online) (a) Crystal structure of 2H-TaSe$_2$, band structure
 of (b) 2H-TaSe$_2$ and (c) 2H-TaSeS along $\Gamma -M-k-\Gamma$ direction. 
Density of states and Fermi surface of (d) and (f) 2H-TaSe$_2$ and (e) and (g)
 2H-TaSeS. Slight modification in the band structure can be observed due to 
doping which can be taken as an effect of disorder due to doping.}
\label{fig5}
\end{center}
\end{figure*}
%---------------------------------------------
%---------------------------------------------
\begin{figure*}[h]
\begin{center}
\epsfig{file=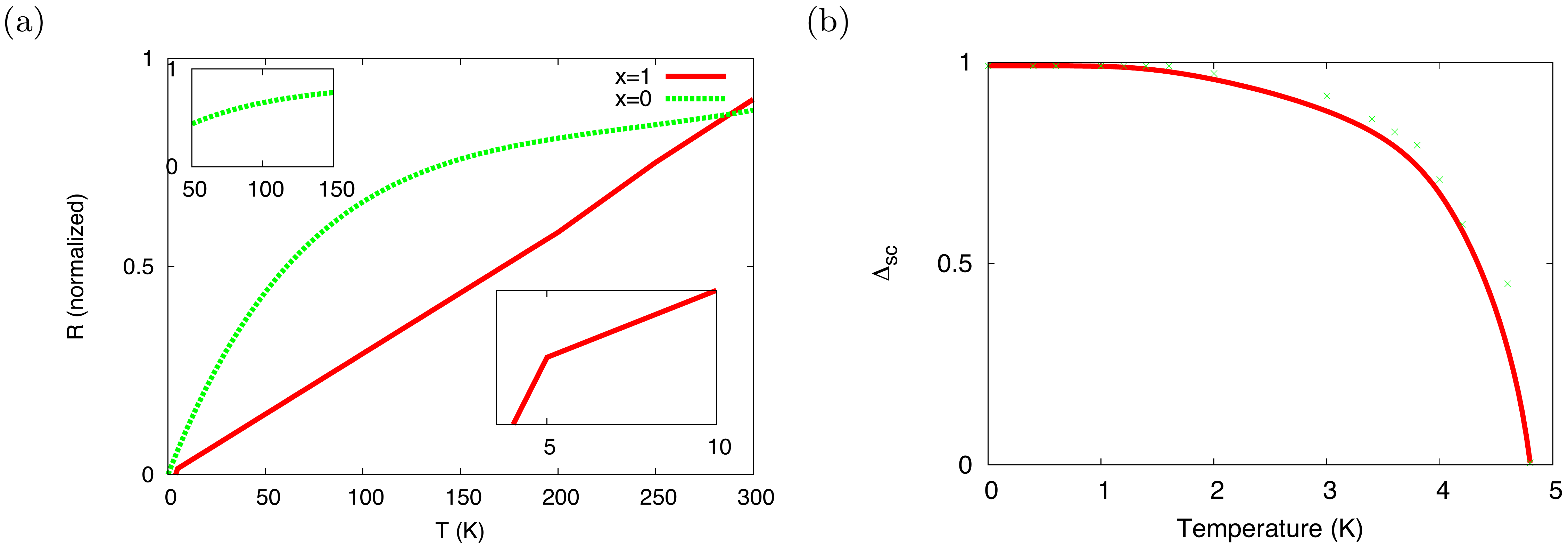,trim=0.0in 0.0in 0.0in 0.0in,clip=false, width=160mm} 
\caption{(Color Online) (a)Main panel shows DMFT resistivity with temperature 
for x=0 and x=1. Upper inset presents resistivity with parent 2H-TaSe$_2$ 
around 120 K while lower inset presents resistivity of 2H-TaSeS around 5~K. 
(b) DMFT superconducting order parameter plot with temperature for 
2H-TaSeS: green points represent actual date points while the red line is a fit.}
\label{fig6}
\end{center}
\end{figure*}
%---------------------------------------------
\noindent For better understanding we provided here spatial distribution of CDW order and superconducting gap  with different disorder strengths in both random and clustered disorder configurations in Fig.~1. The spatial profiles of disorder ((a), (d), (g), (j)),  $\Delta_{SC}$ ((b), (e), (h), (k)) and $\Delta_{CDW}$ ((c), (f), (i), (l)) are shown in Fig.~1. Though the disorder is completely uncorrelated from site to 
site, emergence of superconducting islands increases with increasing disorder. 
These superconducting islands are seperated from one another by very small CDW regions and the size of the islands is the measure of coherence length and can be tuned 
with disorder strength and superconducting pair potential. 
For initial assumption we considered here incommensurate CDW and start from an initial guess of CDW and superconducting order. We find that: 

(i) random potential fluctuations which are not spatially coherent cannot lead to the reappearance of superconductivity. 

(ii) there is a connection between the coherence length of superconductivity and the wavelength of CDW vectors.

(iii) we have considered disorder on a microscopic scale and the combination of 
the pairing interaction, disorder concentration and amplitude of the impurities. This combination 
leads to superconducting islands separated by insulating region. Our results on the disorder strength dependence of superconducting and CDW order for the two 
types of disorder are shown in Fig.2. For both randomly distributed and clustered disorder, CDW order parameter decreases slowly while the SC order parameter increases, indicating that the emergence of superconducting order takes place in a highly-disordered situation. For both kinds of disorder enhancement of superconductivity may be related to disorder induced changes in stoicheometry and lattice bond lengths. In our model, a disorder term essentially describes a random fluctuation in the chemical potential. In reality, however, such potential fluctuations could appear from any change in the stoichiometry and/or bond lengths.  

However a clustering of impurities is the most probable scenario for a macroscopic superconducting order to appear. For impurities distributed randomly over the two-dimensional space, the reappearance of robust superconducting order will require large concentration of impurities (more than 20 percent). 

(iv) The spectral gap (Fig.3) in one-particle density of states persists through all the ranges of disorder in spite of growing number of sites. Not only does the spectral gap persist in the disordered state, it increases with increasing disorder.

%\begin{figure*}
%(a)
%\includegraphics[angle=0,width=0.6\columnwidth]{dos20.ps}
%(b)
%\includegraphics[angle=0,width=0.6\columnwidth]{dos1.ps}
%(c)
%\includegraphics[angle=0,width=0.6\columnwidth]{dos10.ps}
%\caption{(Color Online) Density of states averaged over all sites and disorder 
%at different clustered disorder 
%strength (a) 2, (b) 6, (c)10 and disorder concentration $40\%$. spectral gap is evident throughout all the disorder strength and it increases with the strength.}
%\label{fig3}
%\end{figure*}

%\begin{figure}
%\includegraphics[angle=0,width=0.8\columnwidth]{CDW_Patterns.ps}
%\caption{(Color Online) Spatial profile of the carrier density modulation $\xi_i$, revealing the CDW pattern at different Q vectors (a) ($Q_x$, $Q_y$)=(1,1), (b) ($Q_x$, $Q_y$)=(0.75,0.75), (c) ($Q_x$, $Q_y$)=(0.5,0.5) and (a) ($Q_x$, $Q_y$)=(0.25,0.25).}
%\label{fig4}
%\end{figure}
\noindent In the present setting (shown in Fig. 1) the CDW wavelength is small to start with: in Fig. 1(b) for example, the CDW is like a checkerboard pattern. What we find is that if we take longer CDW wavelength, SC is not suppressed much (or by negligible amount) by CDW in the absence of disorder. This is also understandable as the density modulation due to CDW occurs over a large length scale which does not affect SC. On the other hand, if CDW has a small wave vector (i.e., rapidly changing carrier density) SC is strongly affected. The macroscopic superconducting coherence sets in over a coherence length which the short-wavelength CDW militates against. There is, therefore, a connection between the CDW wave vector and SC coherence length. Disorder sets in its own length scale in the problem by localizing charges, disturbing the existing charge-density profile, facilitating SC order to emerge. \\
\noindent We varied the $Q$ values as described in the figure~\ref{fig4}. The 
charge 
density modulation can have larger period and the checkerboard pattern (shown 
in plot (a) in figure below), obtained at ($Q_x$, $Q_y$) = (1,1) is the case that we 
focus in our description. We find that the suppression of the superconductivity 
by the CDW order decreases with increasing CDW period (or decreasing the values 
of $Q_x$ and $Q_y$). With a smaller CDW period (e.g., in plot (a)), the carrier 
density fluctuation is stronger and the lattice sites loose long-range 
superconducting coherence. So a large superconducting coherence length will help the global 
superconductivity for its resilience against the CDW order. 

In the TMD compounds, for example TaS$_2$ or TaSe$_2$, the Ta ions form a triangular lattice. However, the problem that we discuss in this work \textit{i.e.} the competition between the order parameters of superconductivity and charge density wave in the presence of the non-magnetic disorder is a general one and it does not depend explicitly on the underlying lattice geometry. The main observation, \textit{i.e.} the reentrance of superconductivity in the presence of non-magnetic potential fluctuation, that we extract from our analysis of square lattice is expected to be present in a triangular lattice also.

%\begin{figure*}
%(a)
%\includegraphics[angle=0,width=0.45\columnwidth]{crystal.ps}
%(b)
%\includegraphics[angle=0,width=0.7\columnwidth]{tase2.bands.ps}
%(c)
%\includegraphics[angle=0,width=0.7\columnwidth]{dope1.bands.ps}
%(d)
%\includegraphics[angle=270,width=0.7\columnwidth]{dost.ps}
%(e)
%\includegraphics[angle=270,width=0.7\columnwidth]{dos.ps}
%(f)
%\includegraphics[angle=270,width=0.5\columnwidth]{tase2.eps}
%(g)
%\includegraphics[angle=270,width=0.5\columnwidth]{dope1.eps}
%\caption{(Color Online) (a) Crystal structure of 2H-TaSe$_2$, band structure
 %of (b) 2H-TaSe$_2$ and (c) 2H-TaSeS along $\Gamma -M-k-\Gamma$ direction. 
%Density of states and Fermi surface of (d) and (f) 2H-TaSe$_2$ and (e) and (g)
% 2H-TaSeS. Slight modification in the band structure can be observed due to 
%doping which can be taken as an effect of disorder due to doping.}
%\label{fig5}
%\end{figure*}

\noindent \textit{DFT+DMFT Results:} As discused above, the normal state of TMDs are dominated by incoherent scattering of electrons condensing in to a more coherent CDW state at $T < T_{CDW}$.  To study the competition between superconductivity and CDW order, we have used DFT+DMFT calculation on the transition metal dichalcogenide 2H-TaSe$_2$, where Se is substituted gradually by S, as reported by Li, et al.~\cite{li_npj}. A random concentration of substitution by this method entails huge computational efforts. We work therefore at only two composition (equal Se and S, $x$=1 and $x$=0). As we show, the chemical substitution does indeed lead to a competition between CDW and SC and our theory captures the essential results of Li, et al. Band structures for 2H-TaSe$_2$ and 2H-TaSe$_{2-x}$S$_x$ have been computed within the WIEN2K. The crystal structure of the unit cell is shown in Fig.~4a. The band structures for both the structures are calculated 
from WIEN2K. Band diagram (Fig.~4b and Fig.~4c) of 2H-TaSe$_2$ shows strong Ta-d$_{z^2}$ character hybridized with Se-p in the two metallic bands crossing the Fermi level due to the 
two layers of TaSe$_2$ coupled by van der Waals interaction\cite{band}. The band diagram of doped TaSe$_2$ presents same two bands crossing the Fermi level with slight modification. The Fermi surface (Fig.5f and Fig.5g) also reveals  the same features. The parent structure shows a Fermi surface with hexagonal symmetry and six hole pockets which remain almost same in the doped structure. The evolution of unit cell parameters with doping is consistent with experimental data~\cite{li_npj}. Since the basic structures of 2H-TaSe$_2$, and 2H-TaSe$_{2-x}$S$_x$ are same, the small changes in parent band structure due to doping can be considered as crystallographic disorder. We will now show how it helps in destroying CDW and stabilizing superconductivity.

%\begin{figure*}
%(a)
%\includegraphics[angle=270,width=0.8\columnwidth]{Fig2.ps}
%(b)
%\includegraphics[angle=270,width=0.8\columnwidth]{ord1.ps}
%\caption{(Color Online) (a)Main panel shows DMFT resistivity with temperature 
%for x=0 and x=1. Upper inset presents resistivity with parent 2H-TaSe$_2$ around 120 K while lower inset presents resistivity of 2H-TaSeS around 5~K. 
%(b) DMFT superconducting order parameter plot with temperature for 2H-TaSeS: green points represent actual date points while the red line is a fit.} 
%\label{fig6}
%\end{figure*}
\noindent The parent compound 2H-TaSe$_2$ has two CDW transitions (120~K and 90~K) down to low temperatures and the reported superconductivity is below 1~K ($\sim$ 0.14~K). It is well known that DMFT is an excellent approximation to compute transport coefficients (directly from DMFT Green's 
function) since irreducible vertex corrections vanish for one-band model and are small for  multiband situations. 
Examination of CDW state through resistivity in this dichalcogenide shows a change in slope of resistivity around 120~K in 2H-TaSe$_2$  (Fig. 6a) which is the onset of incommensurate CDW. The resistivity decreases monotonically thereafter without any feature (Fig.6a inset). The onset of commensurate CDW order at 90~K does not reflect in the resistivity as the condensation of pre-formed excitons already began at a higher temperature~\cite{at2}. We then calculate the resistivity from DMFT (Fig.6a, red lines) for 2H-TaSe$_2$ and found that there is no signature of a CDW for $x=1$ compound. Below 5~K, the resistivity suddenly drops shaply to a very small value indicating the onset of an SC transition at 5~K in the $x$=1 alloy.

\noindent In the excitonic CDW scenario~\cite{at2}, the carrier scattering above T$_{CDW}$ is due predominantly to incoherent pre-formed  excitonic fluctuations in TMD systems.   The CDW transition is argued as a coherence-restoring transition of these pre-formed excitons. In order to discuss a CDW or SC state from a single-site DMFT,we follow the excitonic route~\cite{at2} discussed earlier. An effective model for the incoherent metal is written down and a mean-field analysis for the CDW instability is performed thereon (vide reference [14] and [16]). The partial restoration of coherence in the low temperature phase is achieved through excitonic condensation.  The superconductivity follows in a similar manner when the instability is in the particle-particle channel. The resistivity above the instabilities, tellingly, reflects this excitonic fluctuation with a linear behaviour at high temperature (exponent $n=1$) until CDW or SC order sets in.

Disorder brings in a new length scale in the problem and degrades CDW order by localizing charges and suppressing excitonic fluctuations. However, superconductivity is not affected by disorder, unless it is too strong. This is primarily the  reason why SC shows up at the expense of CDW as disorder increases. This also implies that the resistivity actually goes down with disorder initially (till disorder-induced charge localization comes into effect) in the CDW state, seen in our DMFT calculations. 

\noindent Next we present the superconducting order parameter $\Delta_{SC} \propto \langle c_{a\downarrow} c_{a\uparrow}\rangle$ (shown in Fig.6b) at $x$=1, calculated self-consistently from DMFT, to directly check the SC transition seen in resisivity at 5~K. The order parameter follows $(1-T/T_c)^{1/2}$ behaviour and vanishes near 4.5~K (where the resistivity also drops). Thus both qualititative and quantitative agreement of our theoretical results with earlier experiments strongly support the argument that disorder or doping degrades CDW and facilitates superconducting order. Vanishing of CDW with doping also leads to an increase  in carrier density with doping. So the substitutional disorder stabilizes SC at a higher temperatures than the putative T$_c$. We presented a detailed study of the effect of disorder on the CDW transition, and it appeares that in certain parameter regimes, disorder may facilitate an SC state at the expense of 
charge density order. 

\noindent Tuning from CDW to SC phase can be accomplished by intercalation, doping and 
pressure and has been explored largely through experiments.
The above findings have important indications for SC arising in disordered 
state. We have introduced disorder in the TMD by isoelectronic substitution.
Substituting Se by S increases superconducting T$_c$ and at x=1 there is no sign
 of CDW and resistivity varies linear in T due to disorder induced scattering.
 A finite doping increases carrier density at the Fermi level and disorder 
redistributes electrons and holes in the system, analogous to disordered 
chemical potential which in turn weakens CDW order.
In excitonic CDW, as found earlier in parent system\cite{at2} when excitonic 
fluctuations are maximized CDW is destroyed. In disordered square lattice we 
also found that CDW order is destroyed with increasing disorder strength or 
fluctuation in both clustered and random disorder. These constitute the critical
 collective fluctuations which lead to multiband SC with a finite value of 
superconducting order: this is again in accord with the observation of BdG 
calculation. In conclusion, here we show that disorder induced fluctuation 
reduces CDW and maximizes superconductivity both in BdG calculation and also in TMD alloy.

\section{Acknowledgement}
\noindent SK acknowledges department of science and technology, Govt of India 
women scientist scheme (SR/WOS-A/PM-80/2016(G)) for finance.

\end{document}